# Super-resolution imaging in absolute instruments

*Yangyang Zhou, Zhanlei Hao, Pengfei Zhao and Huanyang Chen*[*]

*Institute of Electromagnetics and Acoustics and Department of Physics, Xiamen University*

**Abstract**:

It has been shown that negative refraction makes a perfect lens. However, with little loss, the imaging functionality will be strongly compromised. Later on, it was proved that positive refraction from Maxwell's fish-eye lens can also makes a perfect lens. However, strong debating happens on the introduced drain problem at the imaging position. In this work, we for the first time find that a solid immersion Maxwell's fish-eye lens could be used for super-resolution imaging. We find that it is due to the perfect focusing and total reflection at the outer interface, such that a super-resolution image is formed at the required position in the air background. This simple mechanism will also be valid for other absolute instruments and more versatile super-imaging systems will be anticipated.

**Main Text:**

Diffraction limit is a fundamental problem that hinders us to observe objects in very details within half wavelength scale [1], owning to high spatial frequency evanescent waves carrying subwavelength information of an object cannot travel to the far field. In 2000, Sir John Pendry discovered that negative index material with $\varepsilon = \mu = -1$ (proposed by Veselago [2]) can enhance and recover the evanescent waves to construct a perfect lens, and break down the diffraction limit [3]. Initiated by the concept of perfect lens, a number of superlenses were demonstrated to project the sub-diffraction-limited imaging at the near field [4,5]. Later, through converting evanescent waves carrying the high spatial frequency information of an object to propagating waves, hyperlenses have been proposed by using metamatrials with a hyperbolic dispersion [6,7] to established the imaging with the assistant of conventional lenses in the far field. Using alternating dielectric and metallic layers in a curved geometry, hyperlenses were designed and manufactured [8, 9]. However, insurmountable manufacturing challenges and huge losses for superlens and hyperlens system is a major obstacle for further applications.

---

[*] kenyon@xmu.edu.cn

In recent years, gradient refractive index lens has been developed rapidly for its excellent capability to control the propagation of electromagnetic waves [10-12] and enable focusing or imaging [13-15] with compact artificially microstructures in the subwavelength scale. As one of them, Maxwell's fish-eye (MFE) lens has drawn much attention, due to its property of perfect imaging in geometrical optics [15,16]. U. Leonhardt found that MFE lens [16] can also demonstrate a perfect lens for wave optics, while an active drain has to be introduced at the image point. There was much debating on this aspect [16-19]. Development of super-resolution imaging about MFE lens has entered the bottleneck. Nevertheless, this lens still demonstrates its power in designing various important devices [20-23] if the evanescent wave components are not that important.

On the other hand, solid immersion lenses (SILs) [24,25] as another alternative to yield super-resolution imaging have been studied extensively, where total reflection happens at the interface to excite evanescent waves and improve the imaging resolution. Different types of SILs have been developed from micrometer-size to nanoscale [26-29]. However, due to the small numerical aperture (NA) that is unable to form a resolved image of an object in weak light and with a bit aberration, the imaging functionality is not satisfied.

In this work, we will combine the solid immersion lens together with MFE lens and other absolute instruments, to achieve a super-resolution imaging system. We will find that MFE lens provides a big NA and at the same time zero aberration. The mechanism of super-resolution imaging is that, the perfect imaging at the interface will induce total reflection due to the refractive index and impedance mismatching, and it is this reflection forms evanescent waves at the lens/air interface for the super-resolution imaging of MFE lens, thereby overcoming the diffraction limit. We will analytically prove that Leonhardt-Maxwell lens [15,16] could be very helpful in super-imaging with the help of solid immersion lens and exploit experiment to verify the effect. However, it is difficult to design and fabricate the sample, as the refractive index profile of the solid immersion MEF lens is complex and drastically change along the radius direction. Considering the universality of solid immersion mechanism for realizing super-resolution imaging, we design and implement a special type of Luneburg lens [30, 31] (another absolute instrument) and verify the super-resolution imaging effect at microwave frequencies.

Let us start with a Maxwell's fish-eye (MFE) lens with a radius R in the air background as shown in Fig. 1(a). It can form perfect imaging in geometric optic [15]. Figure 1(a) describes the refractive index profile and its perfect imaging in geometric optics in the free space. Its refractive index profile varies along the radial direction as follows:

$$n(r) = \begin{cases} \dfrac{2n_0}{1+(r/R)^2} & (0 < r < R) \\ 1 & (r \geq R) \end{cases} \quad (1)$$

where $n_0$ represents the ambient refractive index and $r$ is the distance from the center of the lens. To analyze the imaging performance of the lens in wave optics, we will solve Maxwell's equations with a line current source near the MFE lens. For convenience, we choose cylindrical coordinate as the global coordinate. The electromagnetic (EM) wave is assumed to have transverse electric (TE) polarization. The general wave equation that governs the $Ez$ field can be written as

$$r^2 \frac{d^2 Ez}{dr^2} + r \frac{\partial Ez}{\partial r} + \left(k^2 r^2 - m^2\right) Ez = 0 \quad (2)$$

where $k = nk_0$, $n$ represents the refractive index profile and $k_0$ is the free-space wave number. In the MFE lens, we introduce the wave vector $k = k_0 \left( \dfrac{2n_0}{1+(r/R)^2} \right)$ into the Eq. (2) and the general solution is as following

$$E_z^1 = \sum_{m=-\infty}^{\infty} \left( A_m^1 r^m \left(1+r^2\right)^{\frac{1+\sqrt{1+4k^2}}{2}} H_2 F_1 \left[ \frac{1+\sqrt{1+4k^2}}{2}, \frac{1+\sqrt{1+4k^2}+2m}{2}, 1+m, -r^2 \right] \right) \quad \text{for } 0 \leq r \leq R \quad (3)$$

where $H_2 F_1[a,b,c,z] = \sum_{m=0}^{\infty} \dfrac{a^{(m)} b^{(m)}}{c^{(m)}} \dfrac{z^m}{m!}$ is hypergeometric function and m is an integer number. Similarly, the general solution of the air background can be described as

$$E_z^2 = \sum_{m=-\infty}^{\infty} \left( B_m J_m(k_0 r) + C_m Y_m(k_0 r) \right) e^{im\theta} \quad \text{for } r > R \quad (4).$$

We thereby obtain the whole general solution for Maxwell's equations in the whole space. Based on the above analysis, we use addition theorem [33] to add a line current source as excitation to calculate the imaging effect of the MFE lens analytically (Section 1 in [32]). Considering the experimental implementation, we set R=66mm for the radius of the lens. Generally, a MFE lens with $n_0$ =1 has a matching refractive index at the boundary with respect to the background. Therefore, there is no reflection and evanescent waves arising at the interface of lens and air and it fails to realize super-resolution imaging. To overcome the diffraction limit, we introduce ingeniously total internal reflection (TIR) mechanism exciting evanescent waves at the lens/air interface enlightened by solid immersion lenses. In this case, we choose a solid immersion MFE lens with $n_0$ =2.45. It can focus light rays emitting from one point source to another point perfectly while part of light rays are reflected at the imaging point due to impedance mismatching as shown in Fig. 1(a).

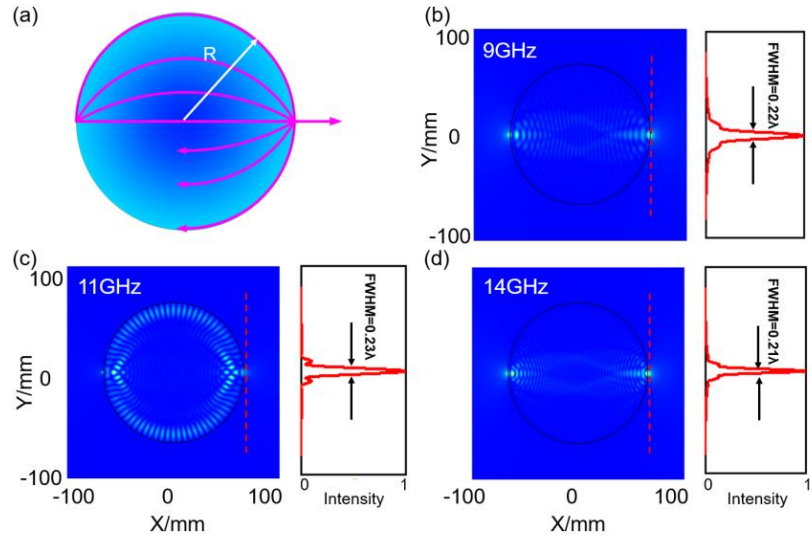

FIG. 1. *Schematic of a solid immersion MEF lens and its imaging functionalities. (a) A gradient refractive index profile and light ray trajectories in the lens. (b-d) Calculated electric field intensity distributions and the corresponding full width at half maxima (FWHM) of the MFE lens at the frequency of 9GH, 11GHz, and 14GHz respectively. The red curves present the normalized electric field intensity along the y-axis direction at the imaging point. The relative FWHM of the imaging point are marked.*

Next, we study the super-resolution imaging of the solid immersion MFE lens for different frequencies. A point source is located at x =−66mm, y=0mm to excite transverse electric (TE) cylindrical waves. Figures 1 (b)-(d) show the electric field intensity patterns and the corresponding FWHM at the frequency of 9GH, 11GHz, and 14GHz respectively. In the figures, the red curves display the electric field intensity and the relative FWHM of the imaging point.

Notably, the corresponding FWHM approaches $0.2\lambda$, which is far below the diffraction limit. Furthermore, we analytically calculate the FWHM for both x-axis direction and y-axis direction at the frequency of 14GHz, as shown in Figure S2 [32]. It demonstrates that the relative FWHM approaches $0.2\lambda$ in both x-axis direction and y-axis direction, which is below the diffraction limit. From the figure, we can see that almost half the super-resolution imaging is in the air background and the relative electric field intensity exponentially decays along the x-axis direction at the imaging point. Overall, the solid immersion MFE lens achieve super-resolution imaging successfully at most frequencies, while for some discrete frequencies, the imaging quality is compromised (see Figure S5 [32]). This is due to the influence of whispering-gallery modes (WGMs) [34], which are specific resonances (or modes) from continuous TIR at the boundary and distribute the imaging. We also verify the phenomenon by analytical calculation (Section 2 in [32]). To analyze the TIR's influence, we investigate the imaging performance of the solid immersion MFE lens by varying the value of $n_0$ from 1 to 4 with a step size of 0.5. The imaging performance about internal reflection ratio is investigated at the frequency of 11GHz. In this process, a point source is located at y =−66 mm, x=0mm to excite TE cylindrical wave. It is clearly seen that a subwavelength spot emerge at the opposite point in the lens and the relative electric field intensity patterns are plotted in Figure S6 [32] for different $n_0$ values. With the increasing of $n_0$, evanescent wave components with larger wavenumber are collected at the imaging point, resulting in the improvement of resolution. The corresponding FWHM satisfies the formula of $0.61/NA$ where $NA=n_0 \sin\theta$ represents the numerical aperture and $\theta$ represents the angles over which the lens can accept or emit light. For the MFE lens, $\theta$ is equal to 90 degree, and FWHM=$0.61\lambda/n_0$. Hence, with the increasing of $n_0$, the resolution of the MFE lens improves as well. However, WGM is inevitably excited owing to the impedance mismatching at the lens/air interface for some $n_0$, which compromises the imaging performance of the lens.

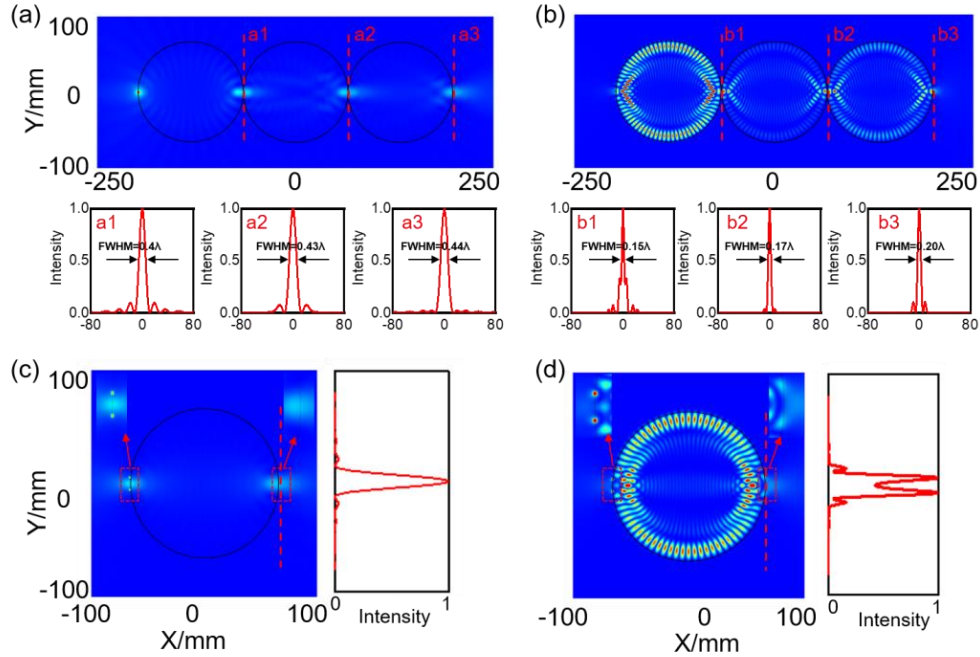

FIG. 2. Comparisons of imaging performance of the classical MFE lens and the solid immersion MFE lens with $n_0$=2.45 at the frequency of 11GHz. (a) The electric field intensity patterns of three interconnected identical classical MFE lenses and the corresponding FWHM at three different imaging positions a1, a2 and a3. (b) The electric field intensity patterns of three interconnected identical solid immersion MFE lenses with $n_0$=2.45 and the corresponding FWHM at three different imaging positions b1, b2 and b3. (c) Imaging performance of the classical MFE lens of which two point sources with a spacing of $0.4\lambda$ is placed at the edge of the lens. The classical MFE lens fails to resolve the two point sources. (d) Imaging performance of the solid immersion MFE lens with $n_0$=2.45 of which two point sources with a spacing of $0.4\lambda$ are placed at the edge of the lens. The two points can be clearly resolved.

To further illustrate the super-resolution imaging capability of the solid immersion MFE lens, we compare the resolution for the classical MFE lens and the solid immersion MFE lens with $n_0$=2.45 in two different ways, as shown in Fig. 2. In the first case, a point source is located at x =−198 mm, y=0mm to excite TE wave at frequency of 11GHz. Figures 2(a) and 2(b) show the $|E|^2$ intensity distribution of a row of MFE lenses consisted of three identical lenses with a radius of R=66mm, for the classical and solid immersion versions respectively. It is noted that the resolution decreases when the imaging point is away from the source point. By comparing those in the classical and solid immersion MFE lenses, we find that the solid immersion MFE lens can maintained the super-resolution imaging even with the imaging point distance increasing as shown in Fig. 2(b). Therefore, we can transfer super-resolution information efficiently using the solid immersion MFE lenses, which is very useful in optical communication systems. In the second case, to verify super-resolution imaging capability of a single solid immersion MFE lens, a pair of identical point sources with a spacing of 0.4λ are placed at the edge of the lens and numerical simulation is performed to analyze results at the frequency of 11GHz, as depicted in Fig. 2(d). It is clearly seen that the two identical point sources are resolved. For comparison, a pair of identical point sources with a spacing of 0.4λ are placed on the edge of the classical MFE lens. Figure 2(c) clearly illustrates that the classical MFE lens failed to distinguish the pair of identical point sources. Based on the above analysis, the proposed solid immersion MFE lens successfully overcome the diffraction limit and achieve the super-resolution imaging.

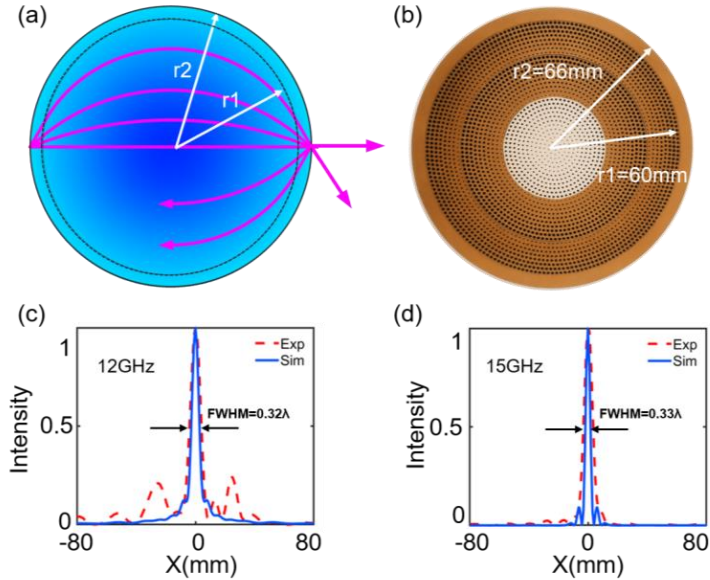

*FIG. 3. Experimental measurements of super-resolution imaging. (a) Schematic of solid immersion general Luneburg (GL) lens, light ray trajectories in the lens, and its gradient index profile. (b) Photograph of the fabricated solid immersion GL lens. (c-d) Comparisons of the normalized electric field intensity profiles along y-axis from -80mm to 80mm of the imaging point between the simulated results and experimental results at the frequencies of 12GHz and 15GHz, respectively. The dotted curves in red represent experimental results and the solid curves in blue represent simulated results, which show good agreement with each other.*

Now we exploit experiments to verify super-resolution imaging of absolute instruments at microwave frequencies. It is shown that a general Luneburg (GL) lens [31] also can form geometric perfect imaging like the above MFE lens, yet with a much easier sample fabrication. Considering the universality of TIR mechanism for realizing super-imaging and easy fabrication, a GL lens is a good alternative to testify the above principle (Section 4 in [32]). In this case, we set the radius of the GL lens r1=60mm, the focal distance $f_1$=$f_2$=r2=1.1*r1=66mm and refractive index $n$=2.45*$n(r)$ to give rise to the refractive index mismatching at the lens/air interface, as shown in Fig. 3(a). The solid immersion GL lens can form perfect imaging in geometric optics. To verify the super-resolution imaging performance of the solid immersion GL lens, full-wave numerical simulations are performed at frequencies ranging from 8GHz to 15GHz. The corresponding results show in Figure S7 [32] and super-resolution imaging ability is demonstrated for the solid immersion GL lens for some discrete frequencies (see in Figure S7 [32]).

To demonstrate the super-resolution performance of the proposed solid immersion GL lens, we fabricate a sample with a radius of 66mm and a height of 10mm, as shown in Fig. 3(b). The lens consisted of two identical samples with a height of 5 mm by drilling air holes of different sizes in different dielectrics with permittivities of 20, 16, 10.2. Based on the effective-medium theory [35], we can calculate the effective permittivity as following

$$\varepsilon_{eff} = f\varepsilon_{air} + (1-f)\varepsilon_d \qquad (5)$$

where $f$ are the volume fraction of the air with a permittivity of $\varepsilon_{air}$, $\varepsilon_{air}$ is the permittivity of substrate dielectric plate and $\varepsilon_{eff}$ is the designed material parameter. The designed details of the effective refractive index and the sizes of unit cells in each layer are shown in Figures S8 and S9, and Table S3 [32]. Full-wave simulation results for the designed lens are shown in Figure S10 [32]. To measure the electric field distribution in the solid immersion GL lens, we utilized a self-made near-field scanning system (see Figure S11 [32]). The detailed test process is depicted in Section 6 in [32]. Figures 3(c) and 3(d) show the simulated and measured normalized electric field intensity of the imaging plane at the frequencies of 12GHz and 15GHz respectively, which demonstrates that the structure can efficiently realize super-resolution imaging. For comparison, it is evident that the experimental results are in good agreement with simulated results, and the effect is valid for a broadband of frequencies (more results see Figure S13 and Table S4 in [32]). The corresponding FWHM is around 0.33 with tiny side-lobes for the experiments. It is clearly seen that the super-resolution imaging performance of the solid immersion GL lens is very good in a broad band of frequencies. Predictably, the proposed solid immersion MFE lens can also overcome the diffraction limit and realize super-resolution imaging.

**Conclusion:**

Enlightened by solid immersion lenses, we introduce ingeniously TIR mechanism exciting evanescent waves at the lens/air interface for the super-resolution imaging of MFE/GL lens. The solid immersion MFE/GL lens could be used to overcome the diffraction limit. We fabricated a sample GL lens using the full-dielectric metamaterials and experimentally verified the super-resolution imaging performance of the solid immersion lens in microwave frequencies. In addition, the effect is robust and valid for a broad band of frequencies. It provides more possibilities for overcoming the diffraction limit from microwave to optical frequencies.

It is possible to pave a way to exciting applications, such as real-time bio-molecular imaging and nanolithography. Furthermore, by combining other methods, such as transformation optics [36, 37] and metasurfaces [38, 39], the MFE lens could be utilized to design many other kinds of absolute instruments (such as rectangle solid immersion MFE lens or perfect hyperlens) for future super-resolution imaging systems.

**Acknowledgment**

This research was Supported by National Natural Science Foundation of China (Grants No. 92050102 and 11874311), National Key Research and Development Program of China (Grant No. 2020YFA0710100), and Fundamental Research Funds for the Central Universities (Grant No. 20720200074). Y.Y.Z. thanks Hao Feng for the experimental assistance.

# Supplementary Materials for
# "Super-resolution imaging in absolute instruments"

*Yangyang Zhou, Zhanlei Hao, Pengfei Zhao and Huanyang Chen\**

*Institute of Electromagnetics and Acoustics and Department of Physics, Xiamen University*

Correspondence author: Huanyang Chen

Email address: kenyon@xmu.edu.cn

## Section 1: Analytical calculation for Maxwell's fish-eye lens

In 1854, the British physicist James Maxwell proposed the famous Maxwell's fish-eye (MFE) lens based on the geometric principle of light: when light passes through MFE lens, it will go around a perfect circle, creating an extraordinary curved light path. Its refractive index profile varies along the radial direction as follows,

$$n(r) = \begin{cases} \dfrac{2n_0}{1+(r/R)^2} & (0 < r < R) \\ 1 & (r \geq R) \end{cases} \tag{S1.1}$$

where $R$ is the radius distance of the lens, r denotes the distance from the center of the lens and $n_0$ represents the ambient refractive index of the lens.

In order to better understand the optical properties of the MFE lens, we study analytically the lens which satisfies the refractive index distribution of the Eq. (S1.1). It is well known that the radial wave equation in the cylindrical coordinate system is in the following form (for transverse electric (TE) polarization),

$$r^2 \frac{d^2 Ez}{dr^2} + r \frac{\partial Ez}{\partial r} + \left(k^2 r^2 - m^2\right) Ez = 0 \tag{S1.2}$$

where $k = nk_0$, $n$ represents refractive index profile, $k_0 = 2\pi/\lambda$ is the free-space wave number and m represents integer number. In the MFE lens, we will take the wave vector $k = 2k_0 n_0 / (1 + (r/R)^2)$ into the Eq. (S1.2) and obtain the general solution of wave equation as follows,

$$E_z = \sum_{m=-\infty}^{\infty} \left( A_m^1 r^m (1+r^2)^{\frac{1+\sqrt{1+4k_0^2 n_0^2}}{2}} H_2 F_1 \left[ \frac{1+\sqrt{1+4k_0^2 n_0^2}}{2}, \frac{1+\sqrt{1+4k_0^2 n_0^2}+2m}{2}, 1+m, -r^2 \right] \right) e^{im\theta} \quad (S1.3)$$

where $H_2 F_1[a,b,c,z] = \sum_{m=0}^{\infty} \frac{a^{(m)} b^{(m)}}{c^{(m)}} \frac{z^m}{m!}$ is the hypergeometric function. The analytical form of the internal MFE lens and external air background can be expressed as follows,

region1: $0 < r \leq R, \{\mu_r, \mu_\theta, \varepsilon_z\} = \left\{ 1, 1, \left( \frac{2n_0}{1+r^2} \right)^2 \right\}$

$$E_z^1 = \sum_{m=-\infty}^{\infty} \left( A_m^1 r^m (1+r^2)^{\frac{1+\sqrt{1+4k_0^2 n_0^2}}{2}} H_2 F_1 \left[ \frac{1+\sqrt{1+4k_0^2 n_0^2}}{2}, \frac{1+\sqrt{1+4k_0^2 n_0^2}+2m}{2}, 1+m, -r^2 \right] \right) e^{im\theta} \quad (S1.4)$$

region2: $R < r, \{\mu_r, \mu_\theta, \varepsilon_z\} = \{1, 1, 1\}$

$$E_z^2 = \sum_{m=-\infty}^{\infty} \left( \alpha_m J_m(k_0 r) + \beta_m Y_m(k_0 r) \right) e^{im\theta} \quad (S1.5)$$

Therefore, we obtain the general solution for the whole region. To further analysis the imaging effect of the MFE lens, we use addition theorem [S1] to introduce a line current source as excitation. Here, we briefly introduce the addition theorem for Hankel function before calculating the electric field distribution inside and outside of the MFE lens, as follows,

$$E_z = -\frac{\omega \mu_0 I_e}{4} \begin{cases} \sum_{m=-\infty}^{\infty} H_m^{(1)}(kr') J_m(kr) e^{im(\theta-\theta')}, r < r' \\ \sum_{m=-\infty}^{\infty} J_m(kr') H_m^{(1)}(kr) e^{im(\theta-\theta')}, r > r' \end{cases} \quad (S1.6)$$

where $w$ is the angular frequency, $\mu_0$ is the vacuum permeability, $I_e$ is the current intensity and $r'$ represent the radius of line current position. Combining the addition theorem for Hankel function with the internal and external analytical form of the MFE lens, the electric field pattern of the MFE lens with a line current source can be easily achieved. If we simply place the line source inside the MFE lens, it will involve the expansion of the hypergeometric functions, which will be very complicated. For simplicity and accuracy, we place the line source in the air background, which is very close to the MFE lens. The addition theorem for Hankel function of cylindrical wave function expansion in homogeneous medium can then be applied. Based on the above consideration, we write the overall analytical solutions in the MFE lens and the air background as follows,

region1: $0 < r \leq R, \{\mu_r, \mu_\theta, \varepsilon_z\} = \left\{1, 1, \left(\dfrac{2n_0}{1+r^2}\right)^2\right\}$

$$E_z^1 = \sum_{m=-\infty}^{\infty} \left( A_m^1 r^m \left(1+r^2\right)^{\frac{1+\sqrt{1+4k_0^2 n_0^2}}{2}} H_2F_1\left[\dfrac{1+\sqrt{1+4k_0^2 n_0^2}}{2}, \dfrac{1+\sqrt{1+4k_0^2 n_0^2}+2m}{2}, 1+m, -r^2\right]\right) e^{im\theta} \quad (S1.7)$$

region2: $R < r \leq r_1, \{\mu_r, \mu_\theta, \varepsilon_z\} = \{1,1,1\}$

$$E_z^2 = \sum_{m=-\infty}^{\infty} \left( B_m^1 H_m^{(1)}(k_0 r) + H_m^{(1)}(k_0 r_1) J_m(k_0 r) \right) e^{im\theta} \quad (S1.8)$$

region3: $r_1 < r, \{\mu_r, \mu_\theta, \varepsilon_z\} = \{1,1,1\}$

$$E_z^3 = \sum_{m=-\infty}^{\infty} \left( C_m^1 H_m^{(1)}(k_0 r) + H_m^{(1)}(k_0 r) J_m(k_0 r_1) \right) e^{im\theta} \quad (S1.9)$$

where $r_1$ is the distance of line current source from the origin and the coefficients $A_m^1, B_m^1, C_m^1$ can be obtained according to the continue boundary conditions of electric fields and magnetic fields ( $E_z^1 = E_z^2$, $H_\theta^1 = H_\theta^2$ ) where $H_\theta = \dfrac{1}{iw\mu_\theta}\dfrac{\partial E_z}{\partial r}$. By applying the continuous boundary conditions of magnetic and electric field at $r = R$, and after some algebra, we obtain the following equations,

$$\begin{cases} A_m^1 2^{\frac{1+\sqrt{1+4k_0^2 n_0^2}}{2}} H_2F_1\left[\frac{1+\sqrt{1+4k_0^2 n_0^2}}{2}, \frac{1+\sqrt{1+4k_0^2 n_0^2}+2m}{2}, 1+m, -1\right] = B_m^1 H_m^{(1)}(k_0) + H_m^{(1)}(k_0 r_1) J_m(k_0) \\ A_m^1 2^{\frac{1+\sqrt{1+4k_0^2 n_0^2}}{2}} \begin{pmatrix} -\left(2k_0 + (m+1)\sqrt{1+4k_0^2 n_0^2}\right) H_2F_1\left[\frac{3+\sqrt{1+4k_0^2 n_0^2}}{2}, \frac{3+\sqrt{1+4k_0^2 n_0^2}+2m}{2}, 1+m, -1\right] \Big/ (1+m) \\ + \left(2m+1+\sqrt{1+4k_0^2 n_0^2}\right) H_2F_1\left[\frac{1+\sqrt{1+4k_0^2 n_0^2}}{2}, \frac{1+\sqrt{1+4k_0^2 n_0^2}+2m}{2}, 1+m, -1\right] \Big/ 2 \end{pmatrix} = \\ k_0 \left( B_m^1 H_m'^{(1)}(k_0) + H_m^{(1)}(k_0 r_1) J_m'(k_0) \right) \end{cases}$$ (S1.10)

By the above equations, we obtain the coefficients $A_m^1$ and $B_m^1$, as follow,

$$\begin{cases} A_m^1 = \dfrac{H_m^{(1)}(k_0 r_1) J_m(k_0) \times k_0 H_m'^{(1)}(k_0) - k_0 H_m^{(1)}(k_0 r_1) J_m'(k_0) \times H_m^{(1)}(k_0)}{\left(k_0 H_m'^{(1)}(k_0) \times \chi_m - H_m^{(1)}(k_0) \times \xi_m\right)} \\ B_m^1 = \dfrac{H_m^{(1)}(k_0 r_1) J_m(k_0) \times \xi_m - k_0 H_m^{(1)}(k_0 r_1) J_m'(k_0) \times \chi_m}{k_0 H_m'^{(1)}(k_0) \times \chi_m - H_m^{(1)}(k_0) \times \xi_m} \end{cases}$$ (S1.11)

where $\xi_m = 2^{\frac{1+\sqrt{1+4k_0^2 n_0^2}}{2}} \begin{pmatrix} -\left(2k_0 + (m+1)\sqrt{1+4k_0^2 n_0^2}\right) H_2F_1\left[\frac{3+\sqrt{1+4k_0^2 n_0^2}}{2}, \frac{3+\sqrt{1+4k_0^2 n_0^2}+2m}{2}, 1+m, -1\right] \Big/ (1+m) \\ + \left(2m+1+\sqrt{1+4k_0^2 n_0^2}\right) H_2F_1\left[\frac{1+\sqrt{1+4k_0^2 n_0^2}}{2}, \frac{1+\sqrt{1+4k_0^2 n_0^2}+2m}{2}, 1+m, -1\right] \Big/ 2 \end{pmatrix}$

and $\chi_m = 2^{\frac{1+\sqrt{1+4k_0^2 n_0^2}}{2}} H_2F_1\left[\frac{1+\sqrt{1+4k_0^2 n_0^2}}{2}, \frac{1+\sqrt{1+4k_0^2 n_0^2}+2m}{2}, 1+m, -1\right]$ . Similarly, the

coefficients $B_m^1$ and $C_m^1$ can be easily derived.

Next, we analyze the solid immersion MFE lens with a radius $R = 1m$ for different $n_0$, where a line current sour source is located at $x = -1.001m$, $y = 0$ to excite TE cylindrical waves (that means in the above formula (S1.8) and (S1.9), we consider that the line source is placed at $r_1 = -1.001$). By analytically calculation, we obtain the internal and external normalized electric field $|E|^2$ of the MFE lens for $n_0 = 1.0$ and $n_0 = 2.5$ as shown in Figure S1 (a) and (b), respectively. In the figure, the relative full width at half maxima (FWHM) are plotted for both horizontal direction for $0 < x < 2m$ and $y = 0$ and vertical direction $x = 1m$ and $-1m < y < 1m$. Figure S1(a) illustrates the normalized field pattern $|E|^2$ in a classical impedance matched MFE lens with $n_0 = 1.0$, and the relative horizontal and vertical FWHM are $0.94\lambda$ and $0.38\lambda$ respectively, which is bigger than the diffraction limit. For the solid immersion MFE with $n_0 = 2.5$, the normalized electric field pattern $|E|^2$ is shown in Figure S1 (b), and the relative horizontal and vertical FWHM of the lens are 0.15λ and 0.19λ respectively, which is far below the diffraction limit. By comparing the FWHM values of the MFE lens with $n_0 = 1.0$ and $n_0 = 2.5$, we can clearly find that the solid immersion MFE can achieve the super-resolution imaging. To further explore the influence of impedance mismatching, we also analytical calculate the relative values of the horizontal and vertical FWHM for the MFE lens from $n_0 = 2.0$ to $n_0 = 3.0$ with a step 0.1 as shown in Table S1. Obviously, the values of FWHM in horizontal and vertical direction vary sharply with $n_0$ increasing when the whispering gallery modes (WGM) [S2] are excited due to continue total reflection. The WGM seriously affects the image quality. As a result, the vertical FWHM is greater than 0.5λ (all details are shown in Table S1). Therefore, we can draw a conclusion that the image resolution of the solid immersion MFE lens in the air background can be significantly improved by the solid immersion mechanism when the WGM are not excited, and the mechanism also works for other perfect geometric lens.

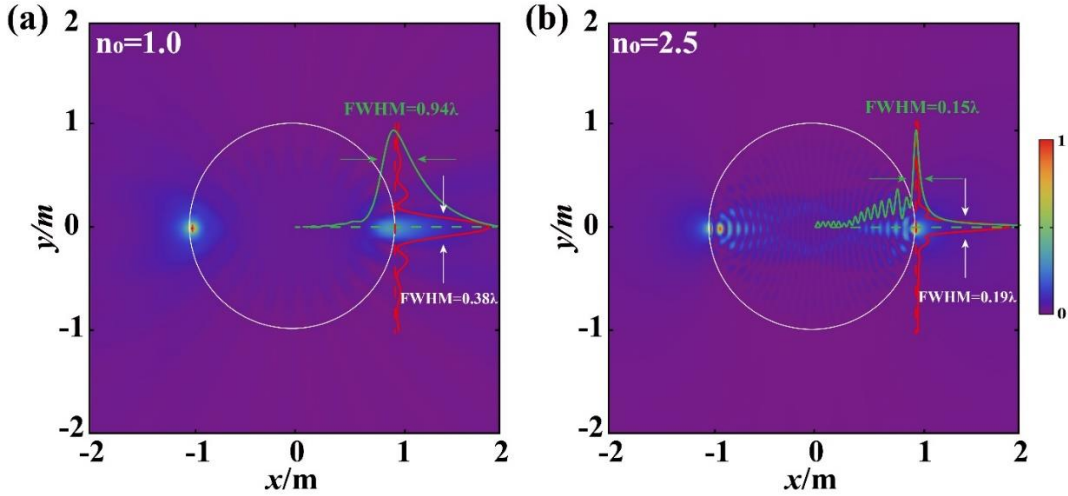

Figure S1. The analytical normalized electric field patterns $|E|^2$ of the MFE lens with (a) $n_0 = 1.0, \lambda=0.5m$, of which the relative $FWHM = 0.94\lambda$ (horizontal direction), $FWHM = 0.38\lambda$ (vertical direction), and (b) $n_0 = 2.5$, $\lambda=0.5m$ of which the relative $FWHM = 0.15\lambda$ (horizontal direction), $FWHM = 0.19\lambda$ (vertical direction).

Table S1 The relative FWHM with different $n_0$

| $n_0$ | 2.0 | 2.1 | 2.2 | 2.3 | 2.4 | 2.5 |
|---|---|---|---|---|---|---|
| **FWHM($x$)** | 0.79λ | 0.19λ | 1.45λ | 1.13λ | 0.67λ | 0.15λ |
| **FWHM($y$)** | 0.22λ | 0.21λ | 0.71λ | 0.74λ | 1.36λ | 0.19λ |
| $n_0$ | 2.6 | 2.7 | 2.8 | 2.9 | 3.0 | |
| **FWHM($x$)** | 1.47λ | 0.97λ | 0.56λ | 0.60λ | 0.22λ | |
| **FWHM($y$)** | 0.82λ | 0.81λ | 1.06λ | 0.18λ | 0.90λ | |

To fully support and demonstrate the super-resolution imaging effect of the solid immersion MFE lens with $n_0 = 2.45$ and $R = 66mm$ in the main text. Here, we analytically calculate the FWHM for both x-axis direction and y-axis direction at the frequency of 14GH (Fig. 1(d) in the main text). A line current source excites TE polarization cylindrical waves and is located at $x = -66.001mm, y = 0$. Figure S2 shows that the normalized electric field intensity distribution $|E|^2$ curve at the imaging point position. The relative horizontal FWHM is below 0.16λ and the normalized electric field intensity $|E|^2$ curve decays exponentially as formula $y = 0.025 + 1.52 \times 10^8 \exp(-x/3.5)$ in the air background, as shown in Figure S1(a). Figure S1(b) describes the vertical normalized electric field intensity $|E|^2$ and the relative FWHM with 0.21λ is marked. Therefore, we prove that the relative FWHM approach 0.2λ in both x-axis direction and y-axis direction, which is below the diffraction limit. The solid immersion MFE achieves successfully super-resolution imaging.

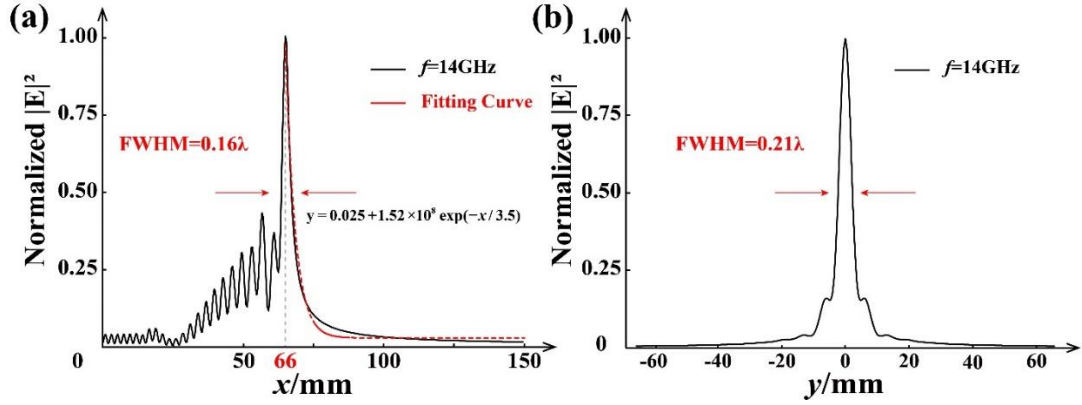

Figure S2. The normalized $|E|^2$ curve of the MFE lens with $n_0 = 2.45$ and $R = 66 mm$ at the frequency of 14GHz for the imaging point. (a) The horizontal normalized $|E|^2$ curve with $FWHM = 0.16\lambda$, (b) The vertical normalized $|E|^2$ curve with $FWHM = 0.21\lambda$.

## Section 2: Analysis for the whispering gallery modes of the solid immersion MFE lens

As we have mentioned above, when the electromagnetic (EM) wave propagates from the solid immersion MFE lens into the air due to continue total internal reflection (TIR) at the boundary, the WGM will be excited in the lens and immensely compromise the imaging quality. In this section, we firstly sovle the WGM of the solid immersion MFE lens with a radius $R=1m$ and $n_0 = 2.0$, and find the corresponding intrinsic wavelengths ranging from 0.2m to 2m in the lens. Later on, we calculate the relative FWHM along the vertical direction of the imaging plane in the solid immerion MFE lens at wavelengths ranging from 0.2m to 2m and find the same FWHM, which is more than $0.38\lambda$ (the corresponding vertical FWHM of the conventional MFE with $n_0 = 1.0$ as shown in Figure S1(a)) at some discrete wavelengthes, where the WGMs are excited. To prove these wavelengths corresponding intrinsic wavelengths of WGMs, we compare the intrinsic wavelength of the excited WGM with the wavelength of the FWHM more than $0.38\lambda$ in the solid immersion MFE lens and find the one-to-one correspondence.

Here, we will analytical calculate the intrinsic wavelength of the excited WGM in the solid immersion MFE lens. By solving two separate 2D Helmholtz equations of the solid immersion MFE lens and the air background, it is not difficult to obtain the solution form as follows

$$\psi_m(r) = \begin{cases} \left( \alpha_m r^m \left(1+r^2\right)^{\frac{1+\sqrt{1+4k_0^2 n_0^2}}{2}} H_2F_1\left[\frac{1+\sqrt{1+4k_0^2 n_0^2}}{2}, \frac{1+\sqrt{1+4k_0^2 n_0^2}+2m}{2}, 1+m, -r^2\right] \right) e^{im\theta} & r \leq R \\ \beta_m H_m^{(1)}(k_0 r) e^{im\theta} & r > R \end{cases} \quad (S1.12)$$

By the continue boundary conditions of eletric field at $r = R$, we can obtain the ratio $\alpha_m/\beta_m$ as follows,

$$\frac{\alpha_m}{\beta_m} = \frac{2^{\frac{1+\sqrt{1+16k_0^2}}{2}} H_2F_1\left[\frac{1+\sqrt{1+16k_0^2}}{2}, \frac{1+\sqrt{1+16k_0^2}+2m}{2}, 1+m, -1\right]}{H_m^{(1)}(k_0)} \quad (S1.13)$$

By continue boundary conditions of magnetic field at $r = R$, the eigen equation which determines the eigenfrequency can be obtained as follows,

$$\frac{\dfrac{-\left(8k^2+(m+1)\left(1+\sqrt{1+16k_0^2}\right)\right)\varsigma_m}{1+m} + \dfrac{\left(2m+1+\left(1+\sqrt{1+16k_0^2}\right)\right)}{2}}{\varsigma_m = H_2F_1\left[\frac{1+\sqrt{1+16k_0^2}}{2}, \frac{1+\sqrt{1+16k_0^2}+2m}{2}, 1+m, -1\right]} = \frac{k_0 H_m'^{(1)}(k_0)}{H_m^{(1)}(k_0)} \quad (S1.14)$$

where $\varsigma_m = H_2F_1\left[\frac{3+\sqrt{1+16k_0^2}}{2}, \frac{3+\sqrt{1+16k_0^2}+2m}{2}, 1+m, -1\right]$ is the hypergeometric function.

The above eigen equation is transcendental equation and can only be solved numerically. The Table S2 shows some calculated intrinsic wavelengths ranging from $0.2m$ to $2m$ with respect to different orders $m$ based on Eqs. (S1.13) and (S1.14). Figure S3 shows the analytical WGM parttern for eginwavelength $\lambda=1.38m$ with respect to the order $m=8$.

Table S2 Calculated eigenwavelength with respect to *m*

| Num | λ | *m* | Num | λ | *M* | Num | λ | *M* |
|---|---|---|---|---|---|---|---|---|
| 1 | 0.275 | 27 | 11 | 0.600 | 16 | 21 | 1.035 | 11 |
| 2 | 0.305 | 39 | 12 | 0.635 | 13 | 22 | 1.055 | 9 |
| 3 | 0.380 | 28 | 13 | 0.670 | 12 | 23 | 1.155 | 8 |
| 4 | 0.395 | 21 | 14 | 0.690 | 17 | 24 | 1.130 | 10 |
| 5 | 0.450 | 21 | 15 | 0.745 | 12 | 25 | 1.245 | 9 |
| 6 | 0.465 | 22 | 16 | 0.785 | 13 | 26 | 1.275 | 7 |
| 7 | 0.485 | 19 | 17 | 0.850 | 10 | 27 | **1.385** | **8** |
| 8 | 0.550 | 16 | 18 | 0.900 | 11 | 28 | 1.425 | 6 |
| 9 | 0.565 | 21 | 19 | 0.955 | 12 | 29 | 1.560 | 7 |
| 10 | 0.575 | 16 | 20 | 0.970 | 10 | 30 | 1.790 | 6 |

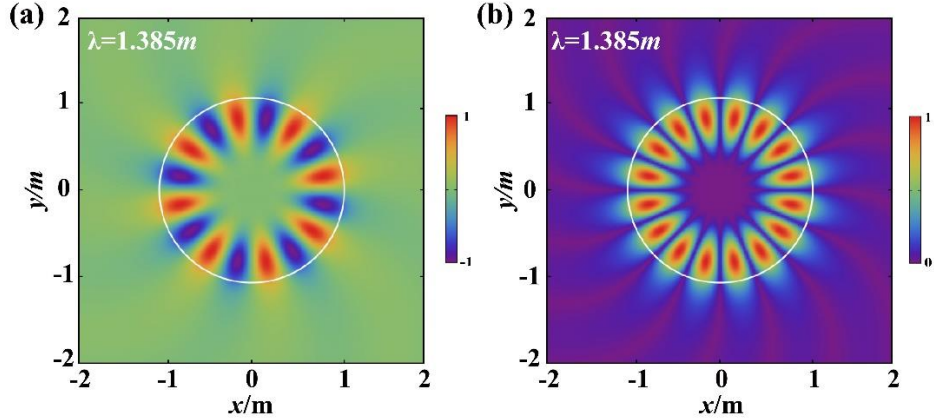

Figure S3. The analytical WGM distribution of the solid immersion MFE lens with $n_0 = 2.0$ at the wavelength $\lambda = 1.38m$. (a) The relative normalized electric field pattern and (b) The normalized electric field intensity pattern $|E|^2$ for the WGM.

In Section 1, we have mentioned that the excited WGM of the solid immersion MFE lens will play a passive role in super-resolution imaging. To prove the disadvantage influence of the WGM, we make a concrete analysis to the resolution of the solid immersion MFE lens with $n_0 = 2.0$ corresponding to the wavelength varying from $0.2m$ to $2m$. The relative

FWHM along vertical direction of the imaging plane are plotted in Figure S4(b). It clearly found that the FWHM of the majority wavelengths are below $0.38\lambda$, while the minority are more than $0.38\lambda$ when the WGMs are excited. By comparing with the Table S2, we prove that the WGMs are excited when the correponding FWHM is mroe than $0.38\lambda$, resulting in a decreasing resolution.Figure S4(a) show the normalized $|E|^2$ distribution. The red curve shows the normalized $|E|^2$ distribution at the left red dashed line at the wavelength $\lambda=0.60m$ and the relative FWHM is more than $0.38\lambda$. We find that the WGM with m=16 is excited in the figure. By comparing with the Table S2, we obtain the intrinsic wavelength of $m=16$. Significantly, the intrinsic wavelength is not unique, e.g., $\lambda=0.465m$ and $\lambda=0.60m$ are eigen wavelengths for $m=16$. We thereby prove that the excited WGMs compromise the super-resolution imaging of the solid immersion MFE lens.

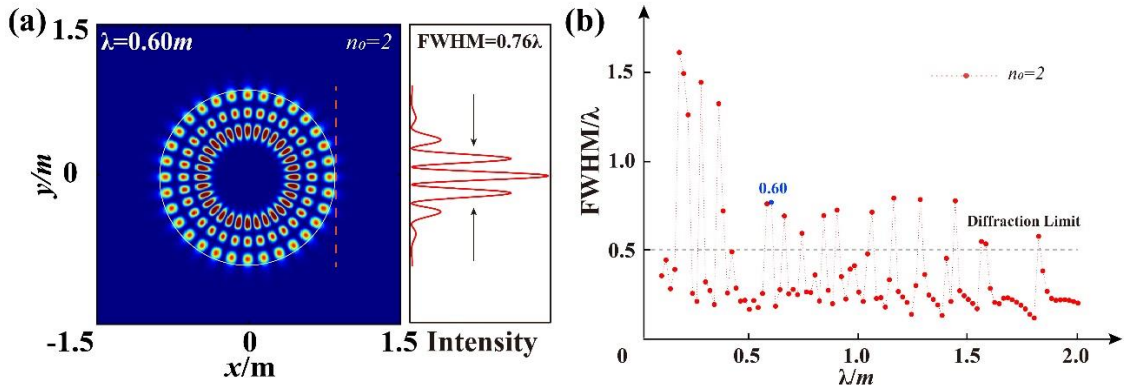

Figure S4. (a) The WGM pattern $|E|^2$ (left) and the red solid curve (right) describes the normalized $|E|^2$ distribution along the left red dashed line for the solid immersion MFE lens with $n_0=2$ at the wavelength $\lambda=0.60m$ (b) The relative FWHM varying with wavelength from $0.2m$ to $2m$.

# Section 3: Numerical simulations of super-resolution imaging performance of the solid immersion Maxwell's fish-eye (MFE) lens and the solid immersion general Luneburg (GL) lens

In this section, we illustrate the super-resolution imaging performance of the solid immersion MFE lens and the solid immersion GL lens. Full-wave simulations are performed with the commercial finite element software COMSOL MULTIPHYSICS. Firstly, we investigate the broadband property of the solid immersion MFE lens with $n_0$=2.45 by varying the frequencies from 8GHz to 15GHz. A point source (line current) is located at x=0mm, y =−66mm (the center of the MFE lens is at the origin) to excite transverse electric (TE) wave (Ez polarization). Figures S5(a)-(h) show the normalized field intensity patterns of the solid immersion MFE lens, where the red curves respectively denote the normalized field intensity at the imaging point. We observe that the corresponding full width at half maximum (FWHM) is less than 0.28λ, which is far below the diffraction limit at the frequency of 8 GHz, 9GHz, 11GHz, 13GHz and 14GHz. Owing to disruption of whispering gallery modes (WGMs), the immersion solid MFE lens fails to achieve super-resolution imaging at the frequency of 10GHz, 12GHz and 15GHz. It is worth noting that WGMs are also excited and slightly distribute the super-resolution imaging in the MFE lens at the frequency of 8GHz, 11GHz and 13GHz.

Secondly, we also investigate the super-resolution imaging performance of the solid immersion MFE lenses by varying the values of $n_0$ at the frequency of 11GHz. Figures S6 (a)-(h) depict the normalized field intensity patterns of the solid immersion MFE lenses, where the red curves respectively display the normalized field intensities. It is noted that with $n_0$ increasing from 1 to 4.5, the TIR phenomenon at the lens/air interface is gradually enhanced. Consequently, evanescent waves with larger wavenumber are excited by TIR at the lens/air interface and participate in imaging improvement. The corresponding FWHM of imaging point decrease from 0.39λ to 0.12λ presenting gradually resolution improvement, as presented in Figures S6 (a)-(h). From figures, we find that WGMs compromise the imaging performance when values of $n_0$ are equal to 2, 2.5 and 4 and the immersion MFE lenses fail to achieve super-resolution imaging.

Thirdly, to verify the super-resolution imaging performance of the solid immersion GL lenses, full-wave simulations are performed to calculate electric field intensity distributions, where a point source (line current) is located at x=0mm, y =−66mm (the center of the GL lens is at the origin) to excite TE wave. To investigate the broadband property of the solid immersion GL lens with $n_0$=2.45, we analyze the imaging performance at the frequencies varying from 8GHz to 15GHz. Relative results are presented in s S7 (a)-(h), where the normalized field patterns of the solid immersion GL lens are shown. The red curves respectively denote the normalized field intensity. We observe that the corresponding FWHM is less than 0.26λ, which is far below the diffraction limit, when WGMs do not affect the imaging of lens at frequency from 10GHz to 15GHz. The solid immersion GL lens can achieve the super-resolution imaging at these frequencies. However, at the frequency of 8GHz and 9GHz, WGMs compromise the super-resolution imaging in the GL lenses.

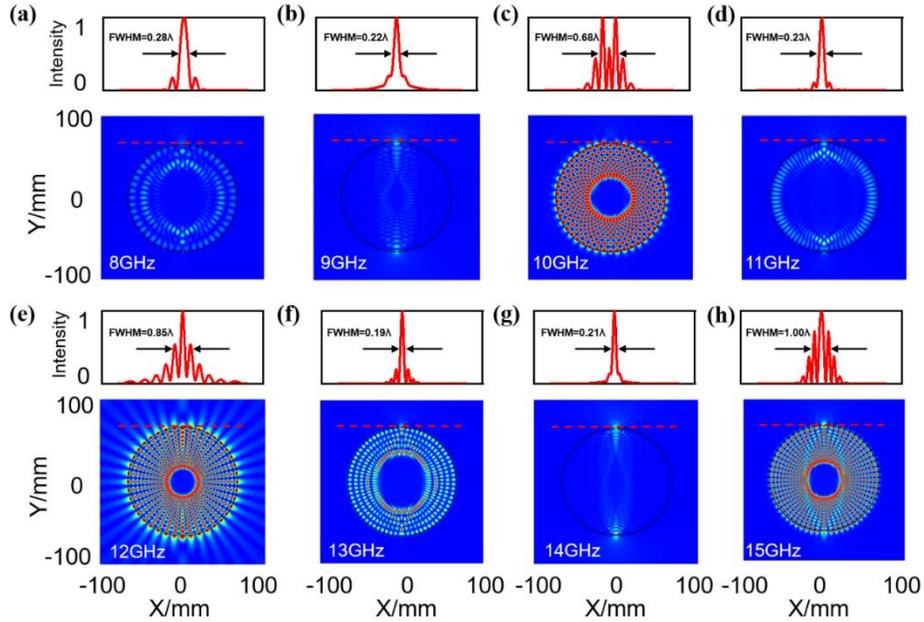

Figure S5. (a-h) Simulated results of the electric field intensity distributions of the solid immersion MFE lens at frequencies ranging from 8GHz to 15GHz. The red curves show the electric field intensity along the x-axis direction from -80mm to 80mm at the position of y =66mm. The FWHM of the imaging point are marked.

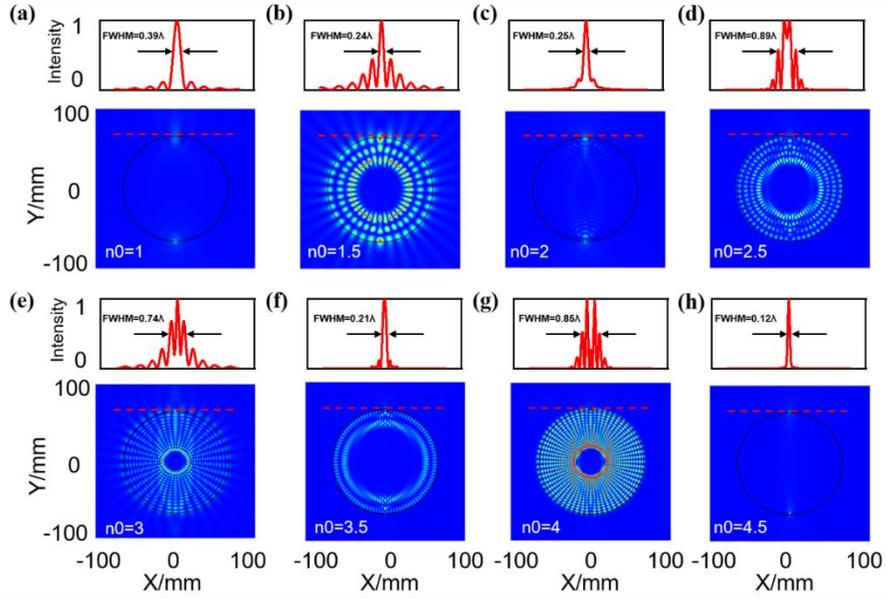

Figure S6. (a-h) Simulated results of the electric field intensity distributions of the solid immersion MEF lens with values of $n_0$ varying from 1 to 4.5. The red curves show the electric field intensity along the x-axis direction from -80mm to 80mm at the position of y =66mm. The relative FWHM of the imaging point are marked.

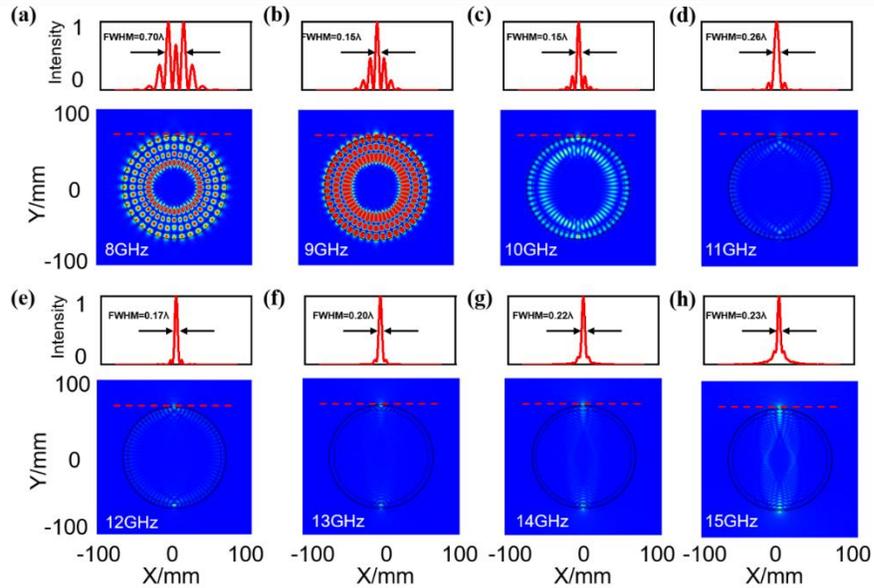

Figure S7. (a-h) Simulated results of the electric field intensity distributions of the solid immersion GL lens at frequencies ranging from 8GHz to 15GHz. The red curves show the electric field intensity along the x-axis direction from -80mm to 80mm at the position of y =66mm. The relative FWHM of the imaging point are marked.

## Section 4: The structural design and data of the solid immersion general Luneburg lens

The GL lens is a variable index, rotationally symmetric structure which can form perfect imaging of between a pair of circles concentric with the center of the lens. The refractive index profile of the GL lens *n(r)* is given by the following equations:

$$n = \begin{cases} \exp[w(\rho, f_1) + w(\rho, f_2)] & (0 \leq \rho \leq 1) \\ 1 & (\rho > 1) \end{cases}, \quad (S1.15)$$

where $\rho = nr$, *r* is radial distance that normalized with respect to the radius of the lens radius, and $f_1$ and $f_2$ are the radii of the concentric circles that are normalized relative to the lens radius with $f_1, f_2 / 1$. The special function *w(p,s)*, defined as

$$w(\rho, f_i) = \frac{1}{\pi} \int_\rho^1 \frac{\arcsin(x/f_i)}{(x^2 - \rho^2)^{1/2}} dx, \quad i = 1, 2. \quad (S1.16)$$

is called the Luneburg integral [24]. In this section, we illustrate structure design and data of solid immersion GL lens based on effective medium theory [S3]. The proposed solid immersion GL lens includes core region with a gradient refractive index and outer uniform layer with uniform refractive index. We divided the lens into four regions, as shown in Figure S9. For uniform outer layer (region 1 in Figure S9) with a refractive index *n*=2.45, dielectric material (TP-2 dielectric with permittivity of 6) is selected to meet the demand of the refractive index distribution. For the core region of lens (regions 2, 3 and 4 in Figure S9), stepped refractive index profile is utilized to fit the curve of the ideal one and illustrate in Figure S8. The core region of lens is divided into 30 layers as shown by the blue line in Figure S8. The related stepped refractive index profile within the lens varies from the refractive index $n_{min}$= 2.45 at the edge to $n_{max}$ = 4.46 at the center. To realize stepped refractive index profile of the core region of lens, we select three different dielectric materials: TP-2 dielectric with permittivities of 20, 16 and 10.5. The materials in regions 2, 3 and 4 are fabricated by drilling different sizes of holes in three pieces of different dielectric plates with permittivities of 10.5, 16 and 20, respectively. We provide all detailed sizes of holes in each region (see in Table S3). The CAD drawing of the whole-designed lens is also presented in Figure S9 and Computer Numerical Control machining technology is utilized to manufacture designed structures of the lens.

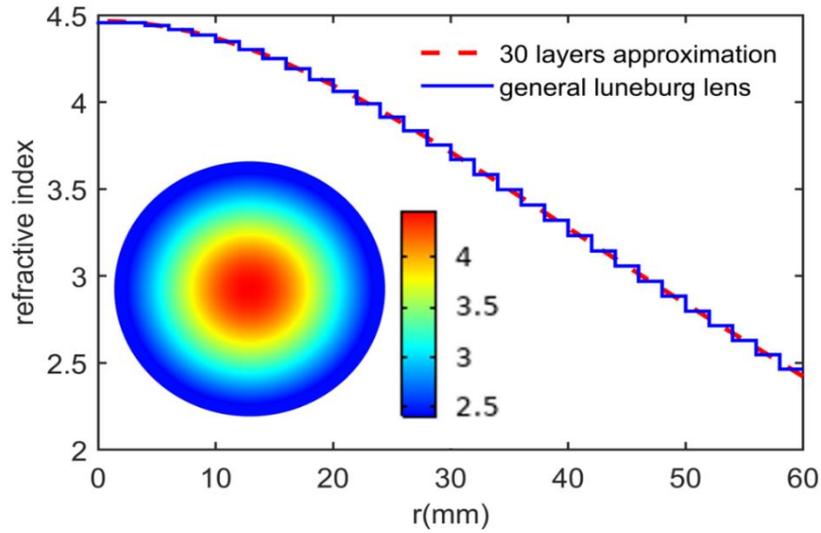

Figure S8. Refractive index profile of the proposed GL's core region (dotted red line) and the discrete (solid blue line) one. The inset describe refractive index distribution of the core region of lens.

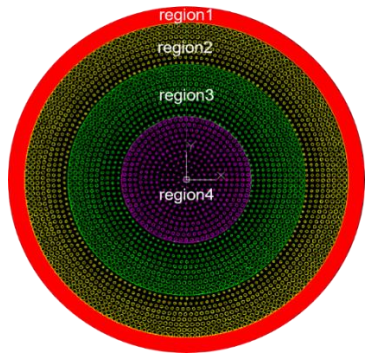

Figure S9. The CAD drawing of the designed GL lens's structures

Table S3. Geometrical parameters of unit cells for different layers in the lens

| layers | Required refractive index | Refractive index realized | Diameters of holes(mm) | Permittivities of the substrate |
|---|---|---|---|---|
| 1 | 2.45 | 2.4494 | no hole | epsilon=6 (region1) |
| 2 | 2.46 | 2.4986 | 1.48 | epsilon=10.2 (region2) |
| 3 | 2.55 | 2.5607 | 1.42 | |
| 4 | 2.63 | 2.6556 | 1.32 | |
| 5 | 2.71 | 2.7406 | 1.22 | |
| 6 | 2.8 | 2.8167 | 1.12 | |
| 7 | 2.88 | 2.9094 | 0.98 | |
| 8 | 2.97 | 2.9875 | 0.84 | |
| 9 | 3.06 | 3.068 | 0.66 | |
| 10 | 3.14 | 3.1452 | 1.44 | epsilon=16 (region3) |
| 11 | 3.23 | 3.2235 | 1.38 | |
| 12 | 3.32 | 3.32 | 1.3 | |
| 13 | 3.41 | 3.4082 | 1.22 | |
| 14 | 3.5 | 3.4888 | 1.14 | |
| 15 | 3.58 | 3.5624 | 1.06 | |
| 16 | 3.67 | 3.6449 | 0.96 | |
| 17 | 3.75 | 3.7311 | 0.84 | |
| 18 | 3.84 | 3.8043 | 0.72 | |
| 19 | 3.91 | 3.8828 | 0.56 | |
| 20 | 3.99 | 3.9558 | 1.08 | epsilon=20 (region4) |
| 21 | 4.06 | 4.0696 | 0.96 | |
| 22 | 4.13 | 4.1365 | 0.88 | |
| 23 | 4.19 | 4.1967 | 0.8 | |
| 24 | 4.25 | 4.2504 | 0.72 | |
| 25 | 4.3 | 4.3194 | 0.6 | |
| 26 | 4.35 | 4.3579 | 0.52 | |
| 27 | 4.39 | 4.3906 | 0.42 | |
| 28 | 4.42 | 4.4115 | 0.38 | |
| 29 | 4.44 | 4.4439 | 0.26 | |
| 30 | 4.46 | 4.4614 | 0.16 | |
| 31 | 4.46 | 4.4614 | 0.16 | |

## Section 5: Numerical verification of the designed structures

To verify validity of the designed structures, full-wave simulations are utilized to examine the electric field intensity distributions which a point source placed at x=0mm, y =−66mm excites the device. As designed structures are rotational symmetric, the unit cells can be considered as an isotropic homogeneous effective medium. We investigate broadband property of the lens by varying the frequency. Figures S10(a)-(h) depict the electric field intensity distributions for the designed lens at frequencies from 8GHz to 15GHz. It is clearly seen that the designed structure can form a tiny finite imaging at a sub-wavelength scale at frequency of 8GHz, 9GHz, 11GHz, 13GHz and 15GHz. At the frequency of 14GHz, due to the influence of excited WGMs, the designed lens faild to achieve super-resolution imaging. By comparions, the simulated results of FWHM are well consistent with the mesured results (see also in Figure S12 and Table S4 in the next section). Due to the influence of the designed gradient periodic structure and WGMs, the designed solid immersion GL lens' FWHM is slightly different from that of the theoretical prediction.

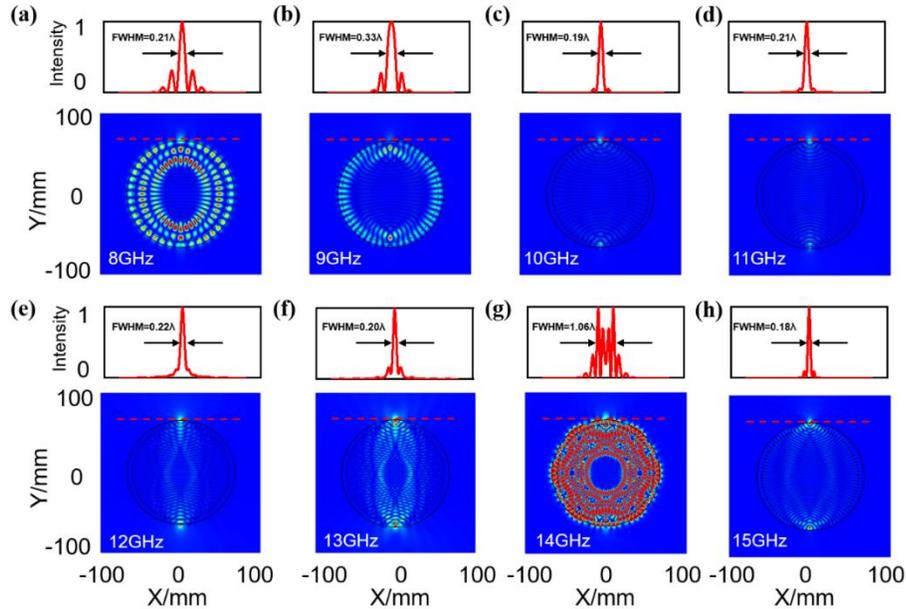

Figure S10. (a-h) Simulated results of the electric field intensity pattern of the designed GL lens with point source frequencies ranging from 8GHz to 15Hz. The red curves represent the electric normolized field intensity along the x-axis from -80mm to 80mm direction at the position of y =66mm. The relative FWHM of the imaging point are marked.

## Section 6: Experimental methods and measured results

For the near field test, the sample is tested in a microwave near-field scanning system as shown in Figure S11(a). The measurement system is consisted of vector network analyzer (Agilent N5230C), three-axis motion controller (Newport ESP301), coaxial probe, two parallel metal plates to measure and set sample. The all details of the measurement system are shown in Figure S11(b). In the process of experiment, the designed solid immersion GL lens sample is fixed on the measuring platform. As point source, a coaxial probe feds the sample and is connected by a coaxial cable from a hole in the measuring platform. To collect the electric field distribution of the designed GL lens, a hole drilled in the parallel metal plate above and a coaxial probe inserted to detect the z-direction component of the electric field. The vector network analyzer is utilized for the excitation and collection of electromagnetic wave signals by control computer.

In the experimental process, the designed solid immersion GL lens is fixed on the measuring platform, as shown in Figure S12(a). A coaxial probe as a feed source is connected by a coaxial cable from a hole in the measuring platform and is close to the edge of the fabricated sample at x=66mm (the center of the sample is located at the origin point (0,0)). The measuring platform with the fixed solid immersion GL lens moves at a spatial resolution of 1mm along the y-axis direction from -80mm to 80mm at the imaging points, and the vector network analyzer collects the signals at each spatial point. We measured the electric field distribution in imaging point along y-axis from -80mm to 80mm at frequencies ranging from 8GHz to 15GHz, as shown in Figure S12. All measured results are given in Figure S13. As comparison, the simulted results also are plotted in Figure S13. The coresponding FWHM of simulted results and experimental results is shown in Table S4. It is clear that the super-resolution imaging behavior occurs in the GL lens at a broad band of frequencies. The experimentally scanning results agree well with numerical simulations.

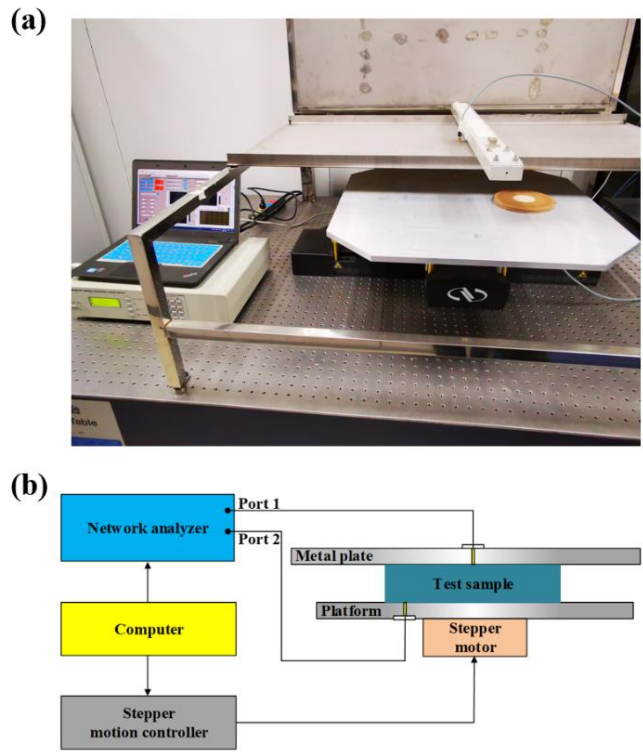

Figure S11. The near field scanning system. (a) The photograph of the near field scanning system. (b) The sketch diagram of the near field scanning system.

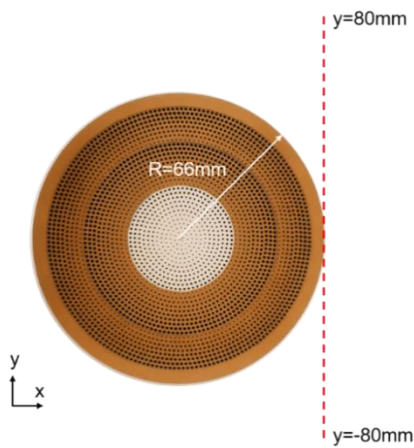

Figure S12. Photograph of fabricated GL lens and scanning track as shown by red dottd line.

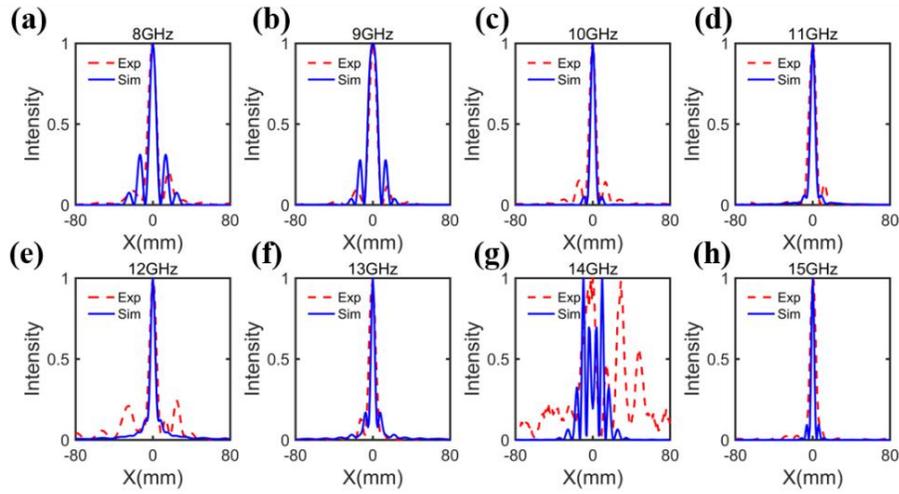

Figure S13. (a-h) Comparisons of the normalized norlized electric field intensity profiles along y-axis from -80mm to 80mm of the imaging point between the simulated resulets and experimental results at the frequencies ranging from 8GHz to 15Hz, respectively. The dotted curves in red represent experimental results and the solid curves in blue represent simulated results. Experimental results show good agreement with simulated results.

Table S4. Comparsions of the FWHM of experimential results and simulated results

| Frequency (GHz) | 8 | 9 | 10 | 11 | 12 | 13 | 14 | 15 |
|---|---|---|---|---|---|---|---|---|
| Simulation's FWHM | $0.21\lambda$ | $0.33\lambda$ | $0.19\lambda$ | $0.21\lambda$ | $0.22\lambda$ | $0.20\lambda$ | $1.06\lambda$ | $0.18\lambda$ |
| Experiment's FWHM | $0.26\lambda$ | $0.27\lambda$ | $0.25\lambda$ | $0.30\lambda$ | $0.32\lambda$ | $0.33\lambda$ | $2.8897\lambda$ | $0.33\lambda$ |